\documentclass{article}
\usepackage{amsmath}
\usepackage{lineno}
\usepackage{arxiv}
\usepackage{tikz}
\usepackage{pgfplots}
\usepackage{graphicx}
\usepackage{algorithm}
\usepackage{algpseudocode}
\usepackage{subcaption}
\usepackage{float}
\usepackage[utf8]{inputenc} 
\usepackage[T1]{fontenc}    

\usepackage{url}            
\usepackage{booktabs}       
\usepackage{amsfonts}       
\usepackage{nicefrac}       
\usepackage{microtype}      
\usepackage{lipsum}
\usepackage{graphicx}
\graphicspath{ {./images/} }
\usepackage{xcolor} %

\pgfplotsset{compat=1.18}

\usepackage{pgfplots}
\usepgfplotslibrary{groupplots}
\pgfplotsset{compat=1.18}
\usepackage[pdftex, colorlinks=true, urlcolor=blue, citecolor=black, anchorcolor=black, linkcolor=black]{hyperref}

\title{Per-Shipment Multi-Agent Reinforcement Learning for Intermodal Freight Routing Under Hurricane Disruption}

\author{
 Aliza Sharmin \\
  Department of Industrial and Systems Engineering\\
  University of Tennessee\\
  Knoxville, TN 37996\\
  \texttt{asharmin@vols.utk.edu} \\
   \And
 Xudong Wang \\
    Department of Industrial and Systems Engineering\\
  University of Tennessee\\
  Knoxville, TN 37996\\
  \texttt{xwang97@vols.utk.edu} \\
  \And
 Mustafa Can Camur \\
  Department of Industrial and Systems Engineering\\
  University of Tennessee\\
  Knoxville, TN 37996\\
  \texttt{mcamur@utk.edu} \\
  \And
 Xueping Li \\
  Department of Industrial and Systems Engineering\\
  University of Tennessee\\
  Knoxville, TN 37996\\
  \texttt{xueping.li@utk.edu} \\
}

\begin{document}
\maketitle
\begin{abstract}
Intermodal freight networks face growing disruption risk from climate extremes that degrade multiple corridors simultaneously. To address this, we formulate freight routing as a Dec-POMDP with per-shipment action granularity and train Independent PPO (IPPO) under Centralized Training with Decentralized Execution, comparing against two heuristic baselines with privileged state access on a 15-hub network under hurricane disruption. Across 30 matched episodes, no single policy dominates: IPPO achieves the highest throughput ($+12.7\%$) and delivery rate while a capacity-aware heuristic leads on Resilience Index (RI) and delay. Under demand surge (2.9:1 capacity ratio), IPPO's RI advantage grows to $+6.4\%$, suggesting learned routing is most valuable when capacity is scarce. A Multi-Agent PPO (MAPPO) variant collapses under train-eval queue mismatch ($\mathrm{RI} = 0.811$); retraining recovers RI to $1.018$ but IPPO still leads on throughput, pointing to residual limitations in centralized critics under per-shipment dispatch.

\end{abstract}


\section{Introduction}

Intermodal freight networks move goods across road, rail, and maritime corridors through consolidation hubs such as ports, rail terminals, and distribution centers. These networks are increasingly exposed to climate-driven disruptions: a single hurricane can reduce port productivity, close highway segments, and constrain rail capacity at the same time, sending ripple effects well beyond the initial point of failure \cite{Sheffi2005,Christopher2004}. As extreme weather events grow more frequent and less predictable, freight systems face disruptions whose timing, severity, and spatial extent are difficult to anticipate \cite{Tang2006SCRisk,Ivanov2020Viable}.

The standard operational response is centralized re-routing or rule-based heuristics \cite{snyder2005reliability}. Both rely on having a global view of the network at decision time, an assumption that often breaks down during active disasters when communication infrastructure is degraded and damage assessments arrive late. Multi-agent reinforcement learning (MARL) under the Centralized Training with Decentralized Execution (CTDE) paradigm \cite{lowe2017multi,foerster2018counterfactual} sidesteps this problem: global information is used during training, but the deployed policy acts on local hub observations alone, matching what dispatchers actually have access to during a crisis. Whether this architectural fit translates into better routing decisions is an empirical question.

A challenge specific to freight is decision granularity. Standard multi-agent formulations assign one action per agent per timestep, but a freight hub may need to dispatch dozens of shipments each step, each headed to a different destination with a different quantity and urgency. Bundling them into a single link selection muddles credit assignment. We instead adopt a per-shipment formulation: each routing decision pairs one shipment with one outgoing link, producing a clean (observation, action, reward) triple. This granularity also raises a question about critic design. Within a single timestep, each dispatch changes the remaining link capacity, so a critic that reads network state only once per step works with an increasingly outdated picture. How much this staleness matters in practice is one of the questions we set out to answer.

A separate challenge is evaluation. The Resilience Index (RI) \cite{Bruneau2003} is widely used to assess infrastructure performance under disruption, but it is computed only after the disruption–recovery cycle is complete and normalizes by each policy's own pre-disruption baseline \cite{hosseini2016review}. Two policies with very different absolute throughput can end up with nearly the same RI if each maintains its own baseline equally well. This is not a flaw in RI: it measures what it is designed to measure; but it means that RI alone can obscure operationally important differences. We therefore pair RI with absolute metrics (total throughput, delivery rate, average delay) to give a fuller picture. 

These considerations lead to our central question: \emph{Under what conditions can decentralized learning agents, making per-shipment routing decisions from local observations, match or outperform hand-crafted heuristics with access to exact network state, and where do simpler approaches suffice?} We investigate this on a 15-hub intermodal network under spatially correlated hurricane disruption, comparing two MARL variants: Independent Proximal Policy Optimization (IPPO) with decentralized critics and Multi-Agent Proximal Policy Optimization (MAPPO) with a centralized critic against two heuristic baselines that read the full network graph.

Our contributions are: (1)~A per-shipment Decentralized Partially Observable Markov Decision Process (Dec-POMDP) formulation \cite{oliehoek2016concise} with destination-conditioned action masking and within-step capacity observations that reflect dispatch-by-dispatch network state changes, (2)~A hurricane disruption model with spatially correlated capacity degradation and stochastic recovery, parameterized from NOAA HURDAT2 historical track statistics \cite{landsea2013atlantic}, (3)~A diagnostic comparison of IPPO and MAPPO under per-shipment dispatch, disentangling the effect of train-eval queue depth mismatch from the within-step observation staleness inherent in centralized critics, (4)~A simulation-based analysis showing that policy effectiveness is capacity-dependent: heuristics perform competitively when spare capacity is abundant, while decentralized learning gains advantage under tighter capacity by mitigating saturation through stochastic load distribution -- a mechanism confirmed by deterministic ablation.

The remainder of the paper is organized as follows. Section~2 reviews related work. Section~3 formulates the problem. Section~4 describes the MARL framework. Section~5 covers experimental setup. Section~6 presents results, discusses findings and limitations. Section~7 concludes the paper.

\section{Literature Review}
This work draws on three research streams: disruption management in freight networks,  multi-agent reinforcement learning (RL) for transportation,  and resilience metrics for learning-based evaluation.

\subsection{Disruption Management in Freight Networks}

Freight networks are vulnerable to disruptions that propagate through interconnected infrastructure, with impacts that extend well beyond the initial failure point \cite{mattsson2015vulnerability,jenelius2006importance}. Research on disruption management has developed along two lines. The \textit{ex ante} stream designs networks to be robust through stochastic programming \cite{Tang2006SCRisk,snyder2005reliability}, robust optimization \cite{Hosseini2019review}, and simulation-optimization hybrids \cite{hruvsovsky2018hybrid}. The \textit{ex post} stream studies how disruptions propagate and how networks recover \cite{ivanov2017simulation}. Both provide tools for network design and post-event analysis, but neither supports the kind of real-time, decentralized decision-making needed during an active disruption \cite{Ivanov2017literature}. A further gap, to our knowledge, is decision granularity: existing models work at the level of aggregate flows or hub-level rules, not individual shipments.

\subsection{Multi-Agent Reinforcement Learning for Transportation Coordination}

The Dec-POMDP framework \cite{oliehoek2016concise} formalizes cooperative decision-making under partial observability. The CTDE paradigm makes this tractable by letting value functions see global state during training while policies act on local observations at test time. Prominent CTDE algorithms include MADDPG \cite{lowe2017multi}, COMA \cite{foerster2018counterfactual}, and QMIX \cite{rashid2020monotonic}.

Proximal Policy Optimization (PPO) \cite{schulman2017proximal} has been extended to multi-agent settings as IPPO (independent critics, local observations only) and MAPPO (shared parameters, centralized critic on global state). Yu et al.\ found that IPPO can match or exceed MAPPO when the global state is redundant or hard for the critic to exploit \cite{yu2022surprising}. Per-shipment dispatch introduces a further complication, the global state is captured once per timestep but must serve as the basis for value estimates across many sequential within-step decisions.

MARL has been applied to traffic signal control \cite{tan2019cooperative} and vehicle dispatch \cite{lin2018efficient}, both settings with one decision per agent per step. Hub-level freight routing is structurally different, agents sit at fixed locations, action spaces are constrained by infrastructure, and multiple shipments must be routed per step. Recent work on intermodal freight resilience has adopted per-shipment re-planning through optimization \cite{filom2025adaptive}, while MARL applications to supply chain disruption have focused on inventory transshipment rather than freight routing \cite{kim2024multi}. Neither, to our knowledge, studies how critic architecture interacts with sub-step decision frequency in a multi-agent routing context.

\subsection{Resilience Metrics and Reward Formulation for MARL}

 The area-based resilience framework introduced by \cite{Bruneau2003} and extended by \cite{cimellaro2010framework} quantifies infrastructure resilience as the integral of performance loss between disruption onset and recovery. While widely adopted for post-event evaluation \cite{ivanov2017simulation,mattsson2015vulnerability}, such metrics are inherently retrospective, as they require observing the full disruption-recovery trajectory and produce only a terminal scalar, providing no intermediate feedback during an episode. In reinforcement learning, this creates a sparse reward problem that slows learning and increases variance \cite{sutton1998reinforcement}. In multi-agent settings, delayed team-level rewards further complicate attribution of individual agent contributions \cite{foerster2018counterfactual}. This motivates shaped per-step rewards that proxy for resilience during training, though alignment between the training signal and terminal evaluation metrics is not guaranteed. Following sections develop the model, training framework, and evaluation protocol targeting these gaps.

\section{PROBLEM FORMULATION}
\label{sec:formulation}

We formulate disruption-aware freight routing as a decentralized sequential decision problem over an intermodal network. The goal is to learn per-shipment routing policies that maintain throughput under spatially correlated disruptions. 

\subsection{Intermodal Network Model}

The freight network is modeled as a directed graph $G = (V,  E)$,  where $V$ is a set of hubs and $E$ is a set of directed links. Each link $\ell \in E$ has a transport mode $m_\ell \in \{\text{road,  rail,  maritime}\}$,  nominal capacity $c_\ell$,  unit cost $c^{\text{cost}}_\ell$,  and travel time $\tau_\ell$. Nodes represent major intermodal transfer points (e.g.,  ports,  rail terminals,  logistics hubs) and links represent aggregated transportation corridors. The set $V$ is partitioned as $V = V^O \cup V^T \cup V^D$,  where $V^O$,  $V^T$,  and $V^D$ denote origin,  transfer,  and destination hubs. Origin hubs generate freight demand,  transfer hubs provide intermediate routing and mode-changes,  and destination hubs are delivery endpoints. Agents are placed at $V^O \cup V^T$, the active decision points. The topology is fixed across episodes; only hurricane parameters vary. 

\subsection{Network Dynamics}

\textbf{Demand Generation:} Freight arrives at origin hubs each timestep via a Poisson process with rate $\bar{\lambda}$,  with quantity $q \sim \mathrm{Uniform}(q_{\min},  q_{\max})$ and a  uniformly random destination.

\textbf{Per-Shipment Dispatch:}
Rather than making one hub-level decision per step, each agent routes every shipment in its queue individually. Each timestep has three phases: (1)~generate demand and reset link capacities; (2)~for each pending shipment, observe the current state, pick a link, and dispatch if capacity allows; decrementing capacity after each dispatch so that later shipments see updated availability; (3)~advance shipments in transit, process deliveries, and apply recovery. A shipment of size $q$ is dispatched on link $\ell$ only if $c^\mathrm{rem}_\ell(t) \geq q$; otherwise it waits and incurs a delay penalty. Under disruption, effective capacity is $c^\mathrm{eff}_\ell(t) = c_\ell (1 - d_\ell(t)) \kappa$ and effective travel time is $\tau^\mathrm{eff}_\ell(t) = \tau_\ell (1 + d_\ell(t))$, where $d_\ell(t) \in [0,1]$ is the disruption severity and $\kappa \in (0,1]$ is a global capacity scaling factor.

\textbf{Destination-conditioned action masking.} To keep
shipments moving toward their destination, the environment restricts
the learned policy's choices to outgoing links that reduce distance
to the target. The raw action index is remapped to a valid link via
modulo indexing, with a fallback to the full action space when no
progress-making link exists. Heuristic baselines do not use this
mechanism; their scoring functions naturally favor geographic progress
but are not strictly constrained to it, allowing them to select
non-progress links when progress-making links lack capacity.

\subsection{Spatially Correlated Disruption Model}

A hurricane event is parameterized by center
$\mathbf{x}^h \in \mathbb{R}^2$ and radius $r$, with
ranges informed by NOAA HURDAT2 statistics
\cite{landsea2013atlantic}. At disruption onset step
$t^h$, each link $\ell$ receives disruption severity
$d_\ell \in [0, 1]$ that decreases linearly with distance
from the storm center ($d_\ell = 0$ outside the footprint). The onset is randomized between steps 30 and 150 of the $H = 300$ episode, so agents must react to observed disruption signals rather than memorizing timing. Recovery is stochastic, each affected link draws an independent recovery rate $\rho_\ell \sim \mathrm{clip}(\mathrm{Exp}(\mu_\rho), \rho_{\min}, \rho_{\max})$,  where $\mu_\rho$ is the mean of the exponential distribution, applied as $d_\ell(t{+}1) = \max(0, d_\ell(t) - \rho_\ell)$.

\subsection{Dec-POMDP Formulation}

The problem is formulated as a Dec-POMDP 
$\langle \mathcal{I},  \mathcal{S},  \{A_i\},  \{O_i\},  T,  R,  \gamma \rangle$,  where $\mathcal{I} = V^O \cup V^T$ is the set of agents,  $\mathcal{S}$ is the global state space,  $A_i$ and $O_i$ are the action and observation spaces for agent $i$,  $T$ is the stochastic transition function,  $R$ is the per-agent reward function,  and $\gamma \in [0, 1)$ is the discount factor.

\smallskip
\noindent\textbf{State space.} The global state $s(t) \in \mathcal{S}$ comprises:
\begin{equation}
  s(t) = \langle\,  \mathbf{q}(t), \; \mathbf{C}(t), \; \mathbf{D}(t), \; F(t) \, \rangle, 
\end{equation}
where $\mathbf{q}(t)$ is the vector of hub inventory levels,  $\mathbf{C}(t)$ the vector of effective link capacities,  $\mathbf{D}(t) \in [0, 1]^{|E|}$ the vector of link disruption severities,  and $F(t)$ the set of in-transit shipment states. The full state is accessible only to the centralized critic during training.

\smallskip
\noindent\textbf{Observation space.} Each per-shipment observation ($d_o = 83$) concatenates hub state ($d_h = 63$), shipment features ($d_s = 3$), and within-step remaining capacities ($K = 17$), as summarized in Table~\ref{tab:obs_layout}. Remaining capacity features update after each dispatch within a step, providing real-time saturation information, a property central to the IPPO vs.\ MAPPO comparison.

\begin{table}[htbp]
\centering
\caption{Observation vector layout ($d_o = 83$). All values normalized
to approximately $[0, 1]$. ``Broadcast'' indicates a network-level
signal; all others are local to the observing hub.}
\label{tab:obs_layout}
\small
\begin{tabular}{rll}
\toprule
\textbf{Dims} & \textbf{Content} & \textbf{Scope} \\
\midrule
\multicolumn{3}{l}{\textit{Hub state ($d_h = 63$)}} \\
3   & Inventory, queue length, timestep        & Local \\
6   & Outgoing capacity \& severity by mode     & Local \\
4   & Incoming in-transit flows                 & Local \\
8   & Neighbor inventory \& queue ($\leq$4)     & Local \\
6   & Disruption context incl.\ RI estimate     & 2 Broadcast \\
34  & Per-link severity \& capacity ($2K$)      & Local \\
2   & Hub identity                              & Local \\
\midrule
\multicolumn{3}{l}{\textit{Shipment features ($d_s = 3$)}} \\
3   & Destination, quantity, waiting time       & Local \\
\midrule
\multicolumn{3}{l}{\textit{Remaining capacity ($K = 17$)}} \\
17  & Remaining cap / full cap per link         & Local \\
\bottomrule
\end{tabular}
\end{table}

\smallskip
\noindent\textbf{Action space \& Transition dynamics.}
Actions select from $K = 17$ outgoing link slots,  remapped through the destination-conditioned mask. Transition dynamics decrement link capacity, advance in-transit flows,  and apply recovery independently of agent actions.

\subsection{Reward Function}
Each dispatch receives a reward combining geographic progress,  a load-balancing penalty,  disruption-dependent scaling,  and a penalty for waiting:
\begin{equation}
R(i, \ell, t)=
\begin{cases}
\bigl[\mathrm{progress}(i, \ell, j)-\alpha_{\mathrm{load}}u_\ell\bigr]\beta_{\mathrm{dis}}, 
& \text{if dispatched}, \\[4pt]
-c_{\mathrm{wait}}, 
& \text{if waiting}.
\end{cases}
\end{equation}
Here,  $i$ denotes the current hub,  $\ell$ the selected outgoing link,  $j$ the shipment's destination hub,  and $t$ the timestep. The progress term:
\begin{equation}
\mathrm{progress}(i, \ell, j)=
\mathrm{clip}\!\left(
\frac{\mathrm{dist}(i, j)-\mathrm{dist}(v_\ell, j)}{d_{\max}}, \, -1, \, 1
\right), 
\end{equation}
where $v_\ell$ is the destination hub of link $\ell$,  $\mathrm{dist}(\cdot, \cdot)$ is Euclidean distance,  and $d_{\max}=L\sqrt{2}$ is the maximum possible distance on the $L\times L$ grid. This term measures normalized geographic progress toward the destination. The destination-conditioned action mask ensures that dispatched shipments always make non-negative progress,  so the lower bound of $-1$ is not reached in practice. The utilization term:
\begin{equation}
u_\ell=\frac{c^\mathrm{eff}_\ell-c^\mathrm{rem}_\ell}{c^\mathrm{eff}_\ell}
\end{equation}
captures link saturation at dispatch,  where $c^\mathrm{eff}_\ell$ is the effective capacity of link $\ell$ under disruption and $c^\mathrm{rem}_\ell$ is the remaining available capacity at the time of dispatch. The multiplier $\beta_{\mathrm{dis}}=1$ under normal conditions and $\beta_{\mathrm{dis}}>1$ during active disruption,  encouraging continued routing,  while $\alpha_{\mathrm{load}}$ controls the strength of the load-balancing penalty and $c_{\mathrm{wait}}$ penalizes delay when capacity is unavailable. Parameter values are given in Section~5.1.

\subsection{Performance Metrics}
We evaluate policies using complementary metrics capturing both relative disruption response and absolute operational performance. Resilience metrics include the Resilience Index (RI), which measures post-disruption throughput maintenance normalized by each policy's own pre-disruption baseline rate (RI $= 1$ indicates sustained baseline; RI $> 1$ indicates recovery exceeding baseline through redundant capacity); recovery time (RT),  the steps from disruption onset until baseline throughput is restored; throughput vulnerability index (TVI),  the maximum single-step throughput drop normalized by baseline rate; and absorption capacity (AC),  the minimum per-step throughput during disruption divided by baseline rate. Operational metrics include delivery rate,  the fraction of generated shipments delivered; average delay,  mean additional time beyond minimum travel time; and total throughput,  cumulative freight delivered over the episode.

\section{MARL FRAMEWORK}
We compare two critic architectures for per-shipment routing under the CTDE paradigm. The question is whether a centralized critic that sees global state, but only once per step, can keep up with the rapid within-step changes caused by sequential dispatches.

\subsection{Policy Architecture}

All $|\mathcal{I}|$ hub agents share a single policy network $\pi_\theta: \mathbb{R}^{d_o} \rightarrow \Delta^K$ that maps the $d_o$-dimensional per-shipment observation to a categorical distribution over $K$ outgoing link slots. The architecture is a three-layer MLP: $d_o \rightarrow d_{\mathrm{hidden}} \rightarrow d_{\mathrm{hidden}} \rightarrow K$. Parameter sharing is appropriate because all hub agents observe structurally homogeneous representations: each hub's observation encodes the same semantic categories (inventory,  outgoing capacities,  disruption signals) regardless of network position. The hub identity feature (the final dimension of the hub observation) allows the shared policy to condition on hub-specific context. Sharing pools experience across all agents simultaneously,  achieving a $|\mathcal{I}|$-fold effective increase in gradient-contributing transitions per episode without additional environment interactions \cite{yu2022surprising}. The same shared policy network is used for both IPPO and MAPPO; the two variants differ only in the critic architecture described below.

\subsection{Value Estimation}

Both critics are used exclusively during training to compute advantage estimates; at execution time the policy is identical in both variants.

\textbf{IPPO:} The critic $V^{\mathrm{IPPO}}_\phi: \mathbb{R}^{d_o} \rightarrow \mathbb{R}$ mirrors the policy architecture ($d_o \rightarrow d_{\mathrm{hidden}} \rightarrow d_{\mathrm{hidden}} \rightarrow 1$,  $\tanh$) and conditions on the same 83-dimensional per-shipment observation,  including within-step remaining capacity features. Because these features update after each dispatch,  every routing decision receives a value estimate reflecting the current link saturation state, not the state at the start of the step. This per-dispatch freshness is structurally important when dozens of shipments are processed sequentially within a single timestep.

\textbf{MAPPO:} The centralized critic $V^{\mathrm{MAPPO}}_\phi: \mathbb{R}^{|\mathcal{I}| \cdot d_h} \rightarrow \mathbb{R}^{|\mathcal{I}|}$ takes concatenated hub observations of all agents ($|\mathcal{I}| \times d_h = 630$ dimensions),  excluding shipment features and remaining capacity,  as it estimates the value of the network state rather than individual dispatches. The architecture applies an input projection followed by ReLU layers: $630 \rightarrow 256 \rightarrow d_{\mathrm{hidden}} \rightarrow d_{\mathrm{hidden}} \rightarrow |\mathcal{I}|$,  producing per-agent value estimates. The global observation is constructed once per timestep before the dispatch loop and reused for all dispatches within that step; the value loss updates only the output head of the acting agent. Because the global observation is fixed within each step,  the centralized critic does not capture within-step capacity changes. 

During training, a per-hub cap of $n_{\mathrm{cap}}$ shipments per step limits queue size, a computational necessity, as uncapped episodes generate ${\sim}150{,}000$ transitions. The cap is removed at evaluation to reflect operational conditions, and queue sizes grow to 50-150 shipments per hub per step. 

\subsection{Policy Optimization}

The policy is trained using PPO \cite{schulman2017proximal} with Generalized Advantage Estimation (GAE) \cite{schulman2015high}. The clipped surrogate objective is:
\begin{equation}
L^{\text{CLIP}}(\theta) = \hat{\mathbb{E}} \left[\min\!\left(
    r(\theta)\, \hat{A}(t), \;\mathrm{clip}\bigl(r(\theta), \,  1{-}\epsilon, \,  1{+}\epsilon\bigr)\, \hat{A}(t)\right)\right], 
\end{equation}

\noindent where $r(\theta) = \pi_\theta(a \mid o) / \pi_{\theta_{\text{old}}}(a \mid o)$ is the importance sampling ratio and $\hat{A}(t)$ is the GAE advantage computed with discount $\gamma$ and trace decay $\lambda_{\text{GAE}}$. An entropy bonus weighted by $c_{\mathrm{ent}}$ encourages exploration. GAE is computed per agent over each agent's temporally ordered shipment decisions within an episode; returns are normalized per episode and advantages per mini-batch. The entropy coefficient is linearly annealed from $c_{\mathrm{ent}}^{\mathrm{start}}$ to $c_{\mathrm{ent}}^{\mathrm{end}}$,  promoting broad exploration early and stable routing with residual stochasticity for load balancing later. Each training episode resets the environment with a per-episode seed controlling demand and hurricane parameters. At each of $H$ timesteps,  demand is generated and agents dispatch shipments via the per-shipment loop (Section~3.2),  subject to the training cap $n_{\mathrm{cap}}$. The hurricane triggers at randomized onset $t^h$. After the episode,  GAE computes per-agent returns and advantages,  and multiple PPO epochs of mini-batch updates are performed with separate gradient clipping for policy and value networks.

\section{EXPERIMENTAL SETUP}
\label{sec:setup}

\subsection{Scenario Configuration}

All experiments use a synthetic 15-hub intermodal freight network on a $100 \times 100$\,km grid: $|V^O| = 5$ origin hubs (west),
$|V^T| = 5$ transfer hubs (interior), $|V^D| = 5$ destination hubs (east), yielding $|\mathcal{I}| = 10$ agents. Distance-based link assignment produces $|E| = 195$ directed links: 58 road ($c_{\mathrm{road}} = 75$), 133 rail ($c_{\mathrm{rail}} = 150$), and 4 maritime ($c_{\mathrm{sea}} = 200$). A global capacity scaling factor $\kappa = 0.7$ is applied, with maximum outgoing degree $K = 17$. Demand arrives at rate $\bar{\lambda} = 50$ shipments per step with quantity $q \sim \mathrm{Uniform}(10, 50)$, over a horizon of $H = 300$ timesteps. At $\kappa = 0.7$, effective link capacities are 52.5 (road), 105 (rail), and 140 (maritime) units, yielding an aggregate capacity-to-demand ratio of approximately 11.7:1. This scale enables controlled diagnosis of critic behavior and routing dynamics under disruption, effects that would be difficult to isolate in larger networks. While the 11.7:1 capacity ratio means that policy differences manifest primarily in throughput and delivery rate rather than resilience index.

\smallskip
\textbf{Hurricane parameters.} Training uses Category~4 hurricanes with per-episode randomization. Onset $t^h$ is randomized (approximately 30-150). Recovery parameters: $\mu_\rho = 0.1$,  $\rho_\ell \in [0.02,  0.30]$,  yielding recovery times from ${\sim}$3 to 45 steps.

\smallskip
\textbf{Training hyperparameters.} Policy: $83 \rightarrow 128 \rightarrow 128 \rightarrow 17$ ($\tanh$). IPPO critic: $83 \rightarrow 128 \rightarrow 128 \rightarrow 1$ ($\tanh$). MAPPO critic: $630 \rightarrow 256 \rightarrow 128 \rightarrow 128 \rightarrow 10$ (ReLU). Learning rates: $3 \times 10^{-4}$. Discount $\gamma = 0.99$; GAE $\lambda_{\text{GAE}} = 0.95$; PPO clip $\epsilon = 0.2$; mini-batch size 512; 4 PPO epochs; 5{, }000 training episodes. Gradient clipping: 0.5 (policy),  1.0 (value). Entropy annealed from $c_{\mathrm{ent}}^{\mathrm{start}} = 0.05$ to $c_{\mathrm{ent}}^{\mathrm{end}} = 0.01$. Training queue cap: $n_{\mathrm{cap}} = 10$ (${\sim}$25{, }000 samples per episode).

\smallskip
\textbf{Reward parameters.} $\alpha_{\mathrm{load}} = 0.3$; $\beta_{\mathrm{dis}} = 1.5$; $c_{\mathrm{wait}} = 0.5$.

\subsection{Evaluation Comparators}

Both baselines read link capacities and disruption severities directly from the network graph,  giving access to exact real-time link states. Learned policies observe these quantities only through the normalized 83-dimensional observation vector. This informational advantage for baselines makes the comparison conservative. Destination-conditioned action masking is applied only to learned policies; heuristic baselines select links directly via their scoring functions with masking disabled,  preserving their intended routing behavior.

\smallskip
\textbf{Re-Route Nearest (RRN).} A greedy heuristic that scores each outgoing link by geographic progress toward the destination,  weighted by available capacity and inversely by disruption severity,  selecting the highest-scoring link. Reactive and deterministic,  given identical hub state and shipment destination, it always selects the same link.

\textbf{Multi-Criteria Heuristic (MCH).} A richer scoring function incorporating mode switching, congestion awareness, downstream viability, and capacity-based link skipping. The scoring logic was iteratively developed with LLM assistance (Claude, Anthropic); the final implementation is a deterministic, stateless, hand-coded function. The heuristic was designed without access to evaluation seeds or results, and its logic is scenario-independent.

\subsection{Evaluation Protocol}

Each policy is evaluated over 30 episodes (seeds 42-71),  ensuring matched disruption scenarios. Learned actions are sampled from the stochastic policy,  consistent with the training distribution. Evaluation processes the full pending queue (no per-hub cap) using the final checkpoint (episode 5{,}000). A demand surge analysis evaluates all policies at demand rates of 100 and 200 shipments per step (30 matched seeds) to examine performance under tighter capacity.

\section{RESULTS AND DISCUSSION}
\label{sec:results}

\subsection{Training Dynamics}

\begin{figure}[!htbp]
  \centering  \includegraphics[width=\textwidth]{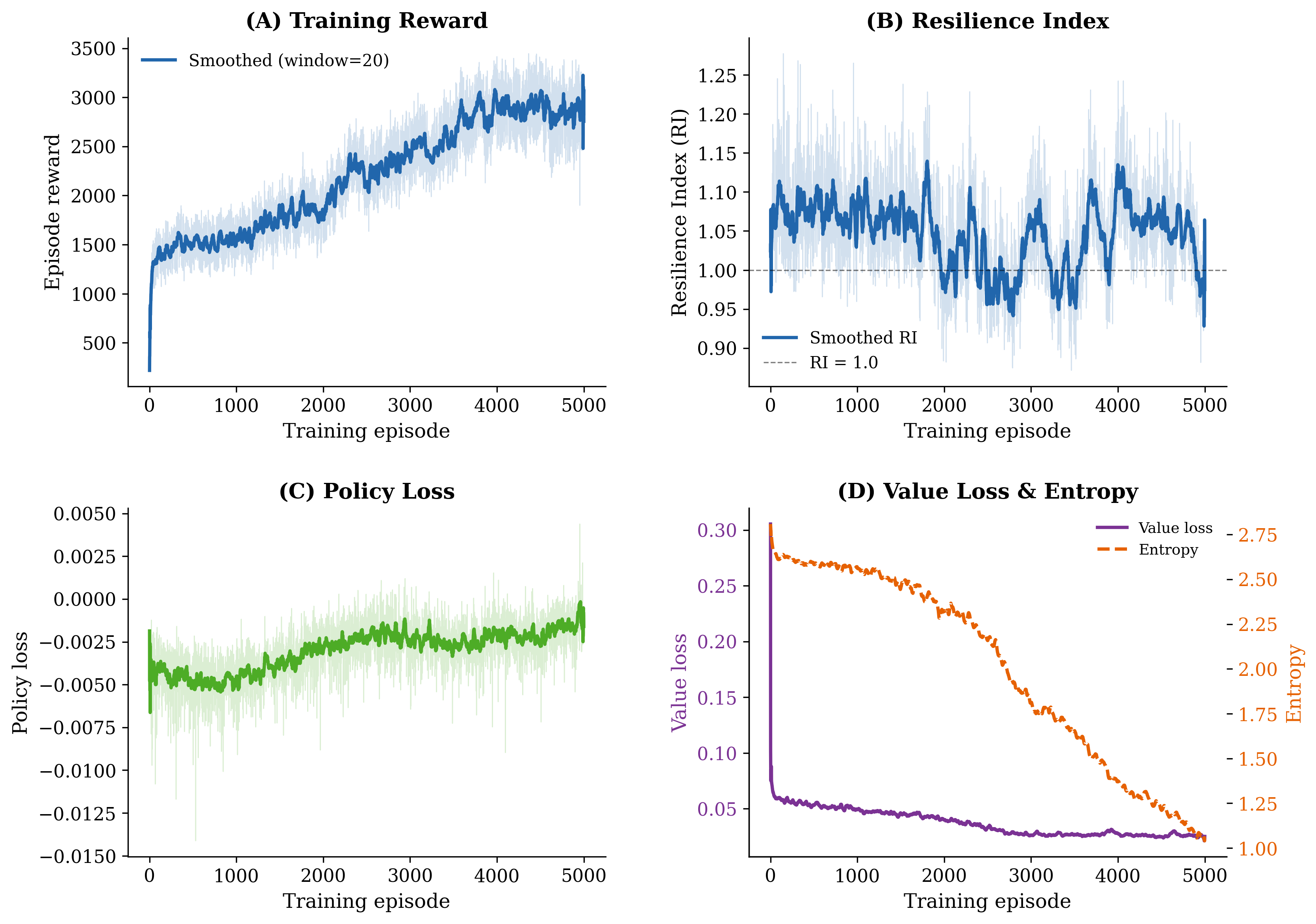}
  \caption{IPPO training progression over 5{,}000 episodes under Category~4 hurricane disruption ($\kappa = 0.7$, 10 agents). (A)~Episode reward converges by episode 3{,}000. (B)~Training RI fluctuates near 1.0. (C)~Policy loss. (D)~Value loss converges rapidly; entropy declines as the policy commits to learned routing patterns.}
  \label{fig:training}
\end{figure}

Figure~\ref{fig:training} shows IPPO training progression over 5{,}000 episodes at $\kappa = 0.7$. IPPO's episode reward converges by roughly episode 3{,}000. Training RI rises sharply in the first 500 episodes and stabilizes slightly below 1.0 (final 100-episode mean: $0.983 \pm 0.033$), indicating that the training policy does not fully maintain its own pre-disruption baseline under the per-hub queue cap. Entropy falls from 2.83 to 1.05 as the policy shifts from broad exploration to committed routing patterns with enough residual randomness for load balancing.

\vspace{-1em}

\subsection{Three-Way Policy Comparison}
\begin{figure}[htbp]
\centering
\begin{tikzpicture}

\definecolor{barblue}{RGB}{76,114,176}
\definecolor{barred}{RGB}{196,108,86}
\definecolor{bargreen}{RGB}{100,168,80}

\pgfplotsset{
    every axis/.append style={
        ybar,
        bar width=14pt,
        ymin=0,
        axis lines*=left,
        axis line style={gray!80},
        tick style={draw=none},
        ymajorgrids=false,
        xmajorgrids=false,
        scaled y ticks=false,
        every axis title/.style={
            font=\footnotesize\bfseries,
            at={(0.5,1.05)},
            anchor=south,
        },
        label style={font=\footnotesize},
        tick label style={font=\footnotesize},
        xtick={1,2,3},
        xticklabels={IPPO, RRN, MCH},
        xmin=0.3, xmax=3.7,
        x tick label style={
            font=\footnotesize,
            anchor=north,
            yshift=-2pt,
        },
        nodes near coords style={
            font=\scriptsize\bfseries,
            anchor=south,
            yshift=1pt,
        },
    },
}

\begin{groupplot}[
    group style={
        group size=3 by 1,
        horizontal sep=1.2cm,
    },
    width=5.4cm,
    height=6cm,
]

\nextgroupplot[
    title={(A) Total Throughput},
    ylabel={Total Throughput (units)},
    ymax=290000,
    ytick={0,50000,100000,150000,200000,250000},
    yticklabels={0,50000,100000,150000,200000,250000},
    y tick label style={
        font=\scriptsize,
        /pgf/number format/fixed,
        /pgf/number format/1000 sep={,},
    },
    nodes near coords={\pgfmathprintnumber[
        fixed, precision=0, 1000 sep={,}
    ]{\pgfplotspointmeta}},
]
\addplot[fill=barblue, draw=none, bar shift=0pt,
    error bars/.cd, y dir=both, y explicit,
    error bar style={black, line width=0.6pt},
    error mark options={rotate=90, mark size=3pt, line width=0.6pt}]
    coordinates {(1, 262090) +- (0,5000)};
\addplot[fill=barred, draw=none, bar shift=0pt,
    error bars/.cd, y dir=both, y explicit,
    error bar style={black, line width=0.6pt},
    error mark options={rotate=90, mark size=3pt, line width=0.6pt}]
    coordinates {(2, 232636) +- (0,5000)};
\addplot[fill=bargreen, draw=none, bar shift=0pt,
    error bars/.cd, y dir=both, y explicit,
    error bar style={black, line width=0.6pt},
    error mark options={rotate=90, mark size=3pt, line width=0.6pt}]
    coordinates {(3, 250266) +- (0,5000)};

\nextgroupplot[
    title={(B) Post-Disruption Throughput Rate},
    ylabel={Post-Disruption Rate (units/step)},
    ymax=980,
    ytick={0,200,400,600,800},
    nodes near coords={\pgfmathprintnumber[fixed,precision=0]{\pgfplotspointmeta}},
]
\addplot[fill=barblue, draw=none, bar shift=0pt,
    error bars/.cd, y dir=both, y explicit,
    error bar style={black, line width=0.6pt},
    error mark options={rotate=90, mark size=3pt, line width=0.6pt}]
    coordinates {(1, 882) +- (0,18)};
\addplot[fill=barred, draw=none, bar shift=0pt,
    error bars/.cd, y dir=both, y explicit,
    error bar style={black, line width=0.6pt},
    error mark options={rotate=90, mark size=3pt, line width=0.6pt}]
    coordinates {(2, 784) +- (0,18)};
\addplot[fill=bargreen, draw=none, bar shift=0pt,
    error bars/.cd, y dir=both, y explicit,
    error bar style={black, line width=0.6pt},
    error mark options={rotate=90, mark size=3pt, line width=0.6pt}]
    coordinates {(3, 845) +- (0,18)};

\nextgroupplot[
    title={(C) Delivery Rate},
    ylabel={Delivery Rate},
    ymax=0.66,
    ytick={0.0,0.1,0.2,0.3,0.4,0.5,0.6},
    nodes near coords={\pgfmathprintnumber[fixed,precision=3]{\pgfplotspointmeta}},
]
\addplot[fill=barblue, draw=none, bar shift=0pt,
    error bars/.cd, y dir=both, y explicit,
    error bar style={black, line width=0.6pt},
    error mark options={rotate=90, mark size=3pt, line width=0.6pt}]
    coordinates {(1, 0.585) +- (0,0.008)};
\addplot[fill=barred, draw=none, bar shift=0pt,
    error bars/.cd, y dir=both, y explicit,
    error bar style={black, line width=0.6pt},
    error mark options={rotate=90, mark size=3pt, line width=0.6pt}]
    coordinates {(2, 0.526) +- (0,0.008)};
\addplot[fill=bargreen, draw=none, bar shift=0pt,
    error bars/.cd, y dir=both, y explicit,
    error bar style={black, line width=0.6pt},
    error mark options={rotate=90, mark size=3pt, line width=0.6pt}]
    coordinates {(3, 0.556) +- (0,0.008)};

\end{groupplot}
\end{tikzpicture}
\caption{Throughput and delivery rate across 30 matched episodes
($\kappa = 0.7$, Cat~4). IPPO: (ours),
RRN: Re-Route Nearest, MCH: Multi-Criteria Heuristic.
Error bars: $\pm 1$ std.}
\label{fig:throughput}
\end{figure}
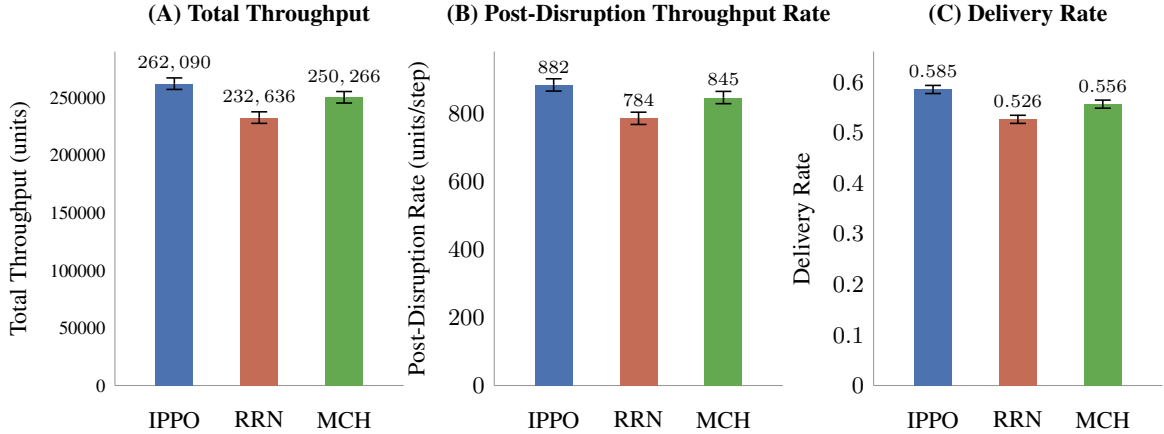

Table~\ref{tab:comparison} and Figure~\ref{fig:throughput} summarize the 30-seed evaluation. The results split cleanly along a throughput-efficiency axis: IPPO moves the most freight, while the MCH routes each shipment most efficiently.

IPPO's throughput of $262{,}090$ units is $12.7\%$ above RRN and $4.7\%$ above MCH. It also achieves the highest delivery rate ($0.585$) and the best disruption-absorption numbers (lowest TVI, highest absorption capacity). This advantage is primarily driven by stochastic load distribution, because the policy samples actions, shipments headed for the same destination are distributed across multiple links, preventing saturation of a single corridor. To isolate this effect, we evaluated the same trained policy with deterministic action selection (argmax over logits). Throughput drops to $208{,}602$ ($-20\%$) and RI falls to $0.833$ (below both baselines) confirming that the stochastic load distribution, not the learned value estimates alone, drives the throughput gain.

The MCH leads on RI ($1.053$) and average delay ($3.3$ steps). Its capacity-skipping logic reads remaining link capacity directly from the network graph and simply waits when preferred links are full, producing very low per-shipment delay at the expense of total volume. RRN, despite having the same privileged state access, concentrates same-destination shipments on one high-scoring link, driving its average delay to $45.5$ steps. Mann-Whitney U tests confirm that RI differences between all policy pairs are not statistically significant (all $p > 0.42$; Mann-Whitney U, two-sided), while throughput and delivery rate differences are highly significant ($p < 0.001$). IPPO beats RRN on RI in only 13 of 30 runs (43\%). The fact that RI cannot distinguish policies differing by $12.7\%$ in throughput underscores the need for the multi-metric evaluation.

\begin{table}[!ht]
\centering
\caption{Policy comparison under Category~4 hurricane ($n = 30$, $\kappa = 0.7$). Best per metric in bold.}
\label{tab:comparison}
\small
\begin{tabular}{lccc}
\toprule
\textbf{Metric} & \textbf{IPPO} & \textbf{RRN} & \textbf{MCH} \\
\midrule
Resilience Index    & $1.044 \pm 0.039$ & $1.043 \pm 0.023$ & $\mathbf{1.053 \pm 0.037}$ \\
Recovery Time       & $\mathbf{9.8 \pm 11.9}$ & $9.8 \pm 10.8$ & $11.7 \pm 14.1$ \\
TVI                 & $\mathbf{0.314 \pm 0.099}$ & $0.322 \pm 0.095$ & $0.347 \pm 0.091$ \\
Absorption Cap.     & $\mathbf{0.686 \pm 0.099}$ & $0.678 \pm 0.095$ & $0.653 \pm 0.091$ \\
Delivery Rate       & $\mathbf{0.585 \pm 0.007}$ & $0.526 \pm 0.005$ & $0.556 \pm 0.003$ \\
Avg Delay (steps)   & $8.9 \pm 0.9$ & $45.5 \pm 1.1$ & $\mathbf{3.3 \pm 0.04}$ \\
Total Throughput    & $\mathbf{262{,}090 \pm 3{,}173}$ & $232{,}636 \pm 1{,}962$ & $250{,}266 \pm 3{,}214$ \\
\bottomrule
\end{tabular}
\end{table}

\vspace{-1em}

\subsection{MAPPO Evaluation}
To investigate how critic architecture interacts with per-shipment dispatch, we trained MAPPO under two conditions: with the standard training queue cap ($n_{\mathrm{cap}} = 10$, matching IPPO) and without it (uncapped, ${\sim}150{,}000$ transitions per episode). Table~\ref{tab:mappo} summarizes the results. With capped training, MAPPO achieves higher training reward than IPPO ($3{,}085$ vs.\ $2{,}907$) but collapses at uncapped evaluation: $\mathrm{RI} = 0.811$, with only 30\% recovery rate. Capping evaluation queues to match training recovers $\mathrm{RI} = 1.030$, identifying the train--eval queue mismatch as the primary cause.

\begin{table}[!htbp]
\centering
\caption{MAPPO diagnostic: disentangling distributional mismatch
from architectural limitation ($\kappa = 0.7$, 30 seeds).
Recovery time computed over recovering runs only.}
\label{tab:mappo}
\small
\begin{tabular}{lccc}
\toprule
& \textbf{MAPPO} & \textbf{MAPPO} & \\
\textbf{Metric} & \textbf{(capped)} & \textbf{(uncapped)} & \textbf{IPPO} \\
\midrule
RI              & $0.811$ & $1.018$ & $\mathbf{1.044}$ \\
Recovery Rate   & $30\%$  & $76.7\%$ & $\mathbf{100\%}$ \\
Throughput      & $211{,}235$ & $201{,}895$ & $\mathbf{262{,}090}$ \\
Delivery Rate   & $0.474$ & $0.454$ & $\mathbf{0.585}$ \\
Recovery Time   & $12.7$ & $45.5$ & $\mathbf{9.8}$ \\
\bottomrule
\end{tabular}
\end{table}

To test whether the failure is purely distributional, we retrained MAPPO without the queue cap. Uncapped-trained MAPPO recovers to $\mathrm{RI} = 1.018$ with 76.7\% recovery rate, above the resilience threshold but still below IPPO on throughput ($201{,}895$ vs.\ $262{,}090$, $-23\%$), delivery rate ($0.454$ vs.\ $0.585$), and recovery time ($45.5$ vs.\ $9.8$ steps). This indicates that the original collapse was primarily distributional, but a residual performance gap persists -- the centralized critic's once-per-step observation still produces less effective routing than IPPO's per-dispatch value estimates, even when training conditions match evaluation.

\subsection{Demand Surge}

\begin{figure}[htbp]
\centering
\begin{tikzpicture}

\definecolor{barblue}{RGB}{76,114,176}
\definecolor{barred}{RGB}{196,108,86}
\definecolor{bargreen}{RGB}{100,168,80}

\pgfplotsset{
    every axis/.append style={
        axis lines*=left,
        axis line style={gray!80},
        tick style={draw=none},
        ymajorgrids=true,
        grid style={dotted, gray!30},
        xmajorgrids=false,
        label style={font=\footnotesize},
        tick label style={font=\footnotesize},
        every axis title/.style={
            font=\footnotesize\bfseries,
            at={(0.5,1.05)},
            anchor=south,
        },
        legend style={
            font=\scriptsize,
            draw=none,
            fill=none,
        },
    },
}

\begin{groupplot}[
    group style={
        group size=2 by 1,
        horizontal sep=1.8cm,
    },
    width=7.5cm,
    height=6cm,
]

\nextgroupplot[
    title={(A) Resilience Index},
    xlabel={Demand rate (shipments/step)},
    ylabel={Resilience Index (RI)},
    xtick={50,100,200},
    xmin=20, xmax=230,
    ymin=0.98, ymax=1.20,
    legend pos=north west,
]
\addplot[black, dotted, thin, forget plot]
    coordinates {(20,1.0)(230,1.0)};
\addplot[barblue, thick, mark=*, mark size=2.5pt,
    error bars/.cd, y dir=both, y explicit,
    error bar style={barblue, line width=0.6pt},
    error mark options={rotate=90, mark size=3pt, line width=0.6pt}]
    coordinates {
        (50, 1.044) +- (0, 0.039)
        (100, 1.074) +- (0, 0.021)
        (200, 1.105) +- (0, 0.020)
    };
\addlegendentry{IPPO}
\addplot[barred, thick, mark=square*, mark size=2.5pt,
    error bars/.cd, y dir=both, y explicit,
    error bar style={barred, line width=0.6pt},
    error mark options={rotate=90, mark size=3pt, line width=0.6pt}]
    coordinates {
        (50, 1.043) +- (0, 0.023)
        (100, 1.054) +- (0, 0.009)
        (200, 1.038) +- (0, 0.008)
    };
\addlegendentry{RRN}
\addplot[bargreen, thick, mark=triangle*, mark size=3pt,
    error bars/.cd, y dir=both, y explicit,
    error bar style={bargreen, line width=0.6pt},
    error mark options={rotate=90, mark size=3pt, line width=0.6pt}]
    coordinates {
        (50, 1.053) +- (0, 0.037)
        (100, 1.149) +- (0, 0.022)
        (200, 1.085) +- (0, 0.019)
    };
\addlegendentry{MCH}

\node[font=\scriptsize\bfseries, barblue, anchor=south west]
    at (axis cs:148,1.115) {+6.4\% RI};
\draw[->, barblue, line width=0.8pt]
    (axis cs:175,1.113) -- (axis cs:198,1.107);

\nextgroupplot[
    title={(B) Total Throughput},
    xlabel={Demand rate (shipments/step)},
    ylabel={Total throughput ($\times 10^3$ units)},
    xtick={50,100,200},
    xmin=20, xmax=230,
    ymin=200, ymax=420,
    scaled y ticks=false,
    y tick label style={
        font=\scriptsize,
        /pgf/number format/fixed,
    },
    legend pos=north west,
]
\addplot[barblue, thick, mark=*, mark size=2.5pt,
    error bars/.cd, y dir=both, y explicit,
    error bar style={barblue, line width=0.6pt},
    error mark options={rotate=90, mark size=3pt, line width=0.6pt}]
    coordinates {
        (50, 262.09) +- (0, 3.17)
        (100, 348.75) +- (0, 4.84)
        (200, 373.29) +- (0, 3.68)
    };
\addlegendentry{IPPO}
\addplot[barred, thick, mark=square*, mark size=2.5pt,
    error bars/.cd, y dir=both, y explicit,
    error bar style={barred, line width=0.6pt},
    error mark options={rotate=90, mark size=3pt, line width=0.6pt}]
    coordinates {
        (50, 232.64) +- (0, 1.96)
        (100, 281.25) +- (0, 0.98)
        (200, 282.06) +- (0, 0.91)
    };
\addlegendentry{RRN}
\addplot[bargreen, thick, mark=triangle*, mark size=3pt,
    error bars/.cd, y dir=both, y explicit,
    error bar style={bargreen, line width=0.6pt},
    error mark options={rotate=90, mark size=3pt, line width=0.6pt}]
    coordinates {
        (50, 250.27) +- (0, 3.21)
        (100, 368.97) +- (0, 2.23)
        (200, 382.19) +- (0, 2.40)
    };
\addlegendentry{MCH}

\node[font=\scriptsize\bfseries, barblue, anchor=south west]
    at (axis cs:140,385) {+32\% vs RRN};
\draw[->, barblue, line width=0.8pt]
    (axis cs:170,383) -- (axis cs:198,375);

\end{groupplot}
\end{tikzpicture}
\caption{Demand surge sensitivity ($\kappa = 0.7$, 30 seeds,
Category~4). (A)~IPPO's RI advantage emerges at $4\times$demand
($+6.4\%$ over RRN). (B)~Throughput: IPPO and MCH
scale; RRN saturates. Error bars:$\pm 1$ std.}
\label{fig:demand_surge}
\end{figure}
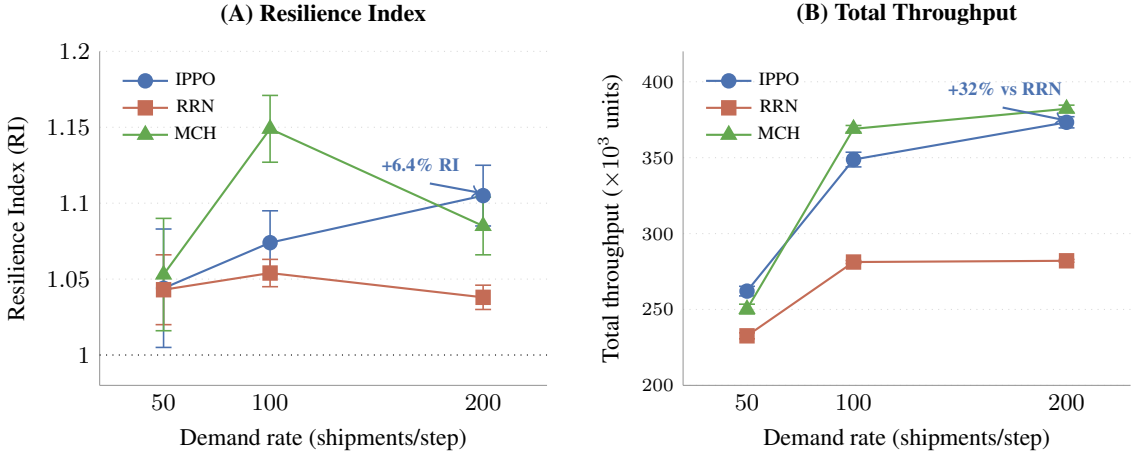

We tested all policies under demand surges of $2\times$ and $4\times$ normal volume (100 and 200 shipments/step) at
$\kappa = 0.7$, using 30 matched seeds, demonstrated in Figure~\ref{fig:demand_surge}. At standard demand (50/step, ratio 11.7:1), the three policies' RI values cluster between $1.043$ and $1.053$. At $2\times$ demand (ratio 5.9:1), the MCH leads on both RI ($1.149$) and throughput ($368{,}971$). At $4\times$ demand (ratio 2.9:1), a crossover appears: IPPO takes the lead on RI ($1.105$ vs.\
$1.085$ for MCH and $1.038$ for RRN), while MCH edges ahead on throughput ($382{,}186$ vs. $373{,}287$). Notably, MCH's RI declines from its $2\times$ peak ($1.149$ to $1.085$), suggesting its capacity-skipping strategy loses effectiveness under extreme load, whereas IPPO's RI rises monotonically with demand pressure. A separate $\kappa$ sweep (training at $\kappa = 0.4$, evaluating across $0.3$-$0.7$ at 5 seeds) tells a consistent story, with the strongest result validated at 30 seeds: at $\kappa = 0.3$ (ratio 5.0:1) IPPO reaches $\mathrm{RI} = 1.281 \pm 0.035$ versus RRN's $1.058 \pm 0.022$ and MCH's $1.019 \pm 0.037$.

\vspace{-1em}

\subsection{When Does Learned Routing Help?}

The practical takeaway is capacity-dependent. When the network has plenty of spare capacity (11.7:1 ratio at standard demand), all three policies perform similarly on RI and the heuristics' direct state access gives them an edge on delay. When capacity tightens, as in the 4x demand surge (ratio 2.9:1), IPPO's stochastic load distribution starts to matter. It prevents the saturation cascades that deterministic policies are prone to, and its RI advantage grows. This pattern (observed on a single synthetic network) suggests a hypothesis worth testing at scale: MARL-based routing may be most valuable for networks operating near capacity limits, which is the condition that arises during major disruptions. 

\subsection{Limitations}
Our 15-hub synthetic network is sufficient as a proof of concept but does not reflect the scale or operational complexity of real freight systems; topology generalization is the most important next step. The per-shipment formulation generates ${\sim}25{,}000$ samples per episode, which may limit scalability. The $\kappa$ sweep results rest on 5 seeds (except $\kappa = 0.3$, validated at 30) and should be treated as exploratory. The informational asymmetry (baselines access exact link states while IPPO sees normalized vectors) makes our comparison conservative but complicates clean attribution. Comparison against optimization-based approaches (e.g., rolling-horizon formulations) would further contextualize the results.

\section{CONCLUSION}
\label{sec:conclusion}

We introduced a per-shipment Dec-POMDP formulation for freight routing under hurricane disruption, where within-step capacity observations enable learned load balancing. On a 15-hub network, no single policy dominates: IPPO achieves the highest throughput ($+12.7\%$) and delivery rate; a capacity-aware heuristic achieves the highest RI and lowest delay. Under demand surge (2.9:1 ratio), IPPO's RI advantage grows to $+6.4\%$, suggesting (pending validation on larger networks) that learned routing may be most valuable when capacity is scarce. Standard MAPPO collapses under train-eval queue mismatch ($\mathrm{RI} = 0.811$); retraining without the mismatch recovers RI to $1.018$ but IPPO still leads, pointing to both distributional sensitivity and residual architectural limitation in centralized critics under per-shipment dispatch. A deterministic ablation confirms that stochastic action sampling, not the learned value estimates alone, drives IPPO's throughput advantage. Future work should test on larger, realistic topologies and explore critic architectures designed for sub-step decision granularity.

\textbf{Acknowledgment}
\label{acknowledgement}
The authors used ChatGPT for language editing.

\textbf{Author Contributions}
\label{author_contributions}

The authors confirm their contribution to the paper as follows: \textit{Study Conception and Design:} all authors (Aliza Sharmin, Xudong Wang , Mustafa Can Camur, Xueping Li); \textit{Data Collection:} all authors; \textit{Analysis and Interpretation of Results:} Aliza Sharmin, Mustafa Can Camur; \textit{Draft Manuscript Preparation:} Aliza Sharmin, Mustafa Can Camur. All authors review the results and approve the final version of the manuscript.

\textbf{Conflict of Interest}
The authors declare that they have no potential conflicts of interest with respect to the research, authorship, and/or publication of this article.

\bibliographystyle{unsrt}  
\bibliography{references}

\end{document}